\documentclass{article}

\begin{document}

\title{
       Teaching Parallel Programming \\
	Using Both High-Level and Low-Level Languages
      }

\author{
Yi Pan\\
Department of Computer Science\\
Georgia State University, Atlanta, GA 30303\\
email: pan@cs.gsu.edu
  }

\maketitle

\begin{abstract}
We discuss the use of both MPI and OpenMP in the teaching of senior
undergraduate and junior graduate classes in parallel programming. 
We briefly introduce the OpenMP standard and discuss why we have chosen to use it in 
parallel programming classes. 
Advantages of using OpenMP
over message passing methods are discussed.
We also include a brief enumeration of some of the drawbacks 
of using OpenMP and how these drawbacks are being addressed by supplementing
OpenMP with additional MPI codes and projects.
Several projects given in my class are also described in this paper.

\end{abstract}

\section{Introduction}

Parallel processing, the method of having many small tasks solve one large problem, 
has emerged as a key enabling technology in modern computing. 
The past several years have witnessed an ever-increasing acceptance and 
adoption of parallel processing, both for high-performance scientific computing 
and for more ``general-purpose'' applications.
The trend was a result of the demand for higher performance, lower cost, 
and sustained productivity. 
The acceptance has been facilitated by two major developments: 
massively parallel processors and the widespread use of
clusters of workstations.

In the last ten years, courses on parallel computing and programming have been
developed and offered in many institutions as a recognition of the
growing significance of this topic in computer science 
\cite{berry, miller, nevison, wilkinson}.  Parallel
computation curricula are still in their infancy, however, and there
is a clear need for communication and cooperation among the faculty
who teach such courses. 

Georgia State University (GSU), like many institutions in the world, 
has offered a parallel programming course at the graduate and Senior undergraduate level 
for several years. Low-level languages and
tools that have been used at GSU for the course included 
Parallel Virtual Machine (PVM) and Message Passing Interface (MPI)
on an SGI Origin 2000 shared memory multiprocessor system.
As we all know, the message passing paradigm has several disadvantages:
the cost of producing a message passing code may be between 5 and 10 times
that of its serial counterpart,
the length of the code grows significantly,
and it is much less readable and less maintainable than the sequential
version.
Most importantly, the code produced
using the message passing paradigm usually
uses much more memory than the corresponding code
produced using high level parallel languages
since a lot of buffer space is needed in the message passing paradigm.
For these reasons,
it is widely agreed that a higher level programming paradigm is essential if parallel systems
are to be widely adopted.
Most schools teaching the course use low-level message passing
standard such as MPI or PVM and have not adopted OpenMP yet 
\cite{berry, miller, nevison, wilkinson}.
To catch up the trend in industries, we decide to teach 
shared-memory parallel programming model besides
the message passing parallel programming model.
This paper describes my experience in using OpenMP as well as MPI to
teach a parallel programming
course at Georgia State University.


\section{About OpenMP}

The rapid and widespread acceptance of shared-memory multiprocessor architectures 
has created a pressing demand for an efficient way to program these systems. 
At the same time, developers of technical and scientific applications in industry and 
in government laboratories find they need to parallelize huge volumes of code in a 
portable fashion. 

The OpenMP Application Program Interface (API)
\cite{openmp}
 supports multi-platform shared-memory parallel programming in C/C++ and Fortran on all architectures, including Unix platforms and Windows NT platforms. Jointly defined by a group of major computer hardware and software vendors, OpenMP is a portable, scalable model that gives shared-memory parallel programmers a simple and flexible interface for developing 
parallel applications for platforms ranging from the desktop to the supercomputer.
It consists of a set of compiler directives and library routines
 that extend FORTRAN, C, and C++ codes to express shared-memory parallelism. 

OpenMP's programming model uses fork-join parallelism: 
master thread spawns a team of threads as needed. Parallelism is added 
incrementally: i.e. the sequential program evolves into a parallel program.
Hence, we do not have to parallelize the whole program at once.
OpenMP is usually used to parallelize loops.
A user finds his most time consuming loops in his code,
and split them up between threads.
In the following, we will give some simple examples to demonstrate
the major features of OpenMP.

Below  is a typical example of big loops in a  sequential C code:

\begin{verbatim}
void main()
{
    double A[100000];
    fo(int i=0;i<100000;i++) {
     big_task(A[i]);
    }
}
\end{verbatim}

In order to parallelize the above code in OpenMP, users just need to
insert some OpenMP directives to tell the compiler how to parallelize
the loop.
A short hand notation that combines the Parallel and work-sharing construct
is shown below:

\begin{verbatim}
void main()
{
       double Res[100000];
#pragma omp parallel for
       for(int i=0;i<100000;i++) 
       {
           big_task(Res[i]);
       }
}
\end{verbatim}

The OpenMP work-sharing for construct basically splits up loop iterations  
among the threads in a team to achieve parallel efficiency. 
By default, there is a barrier at the end of the ``omp for''.
We can use the ``nowait'' clause to turn off the barrier. 

Of course, there are many different OpenMP constructs are available
for us to choose. The major difficult job in parallelize a code using
OpenMP is how to choose an OpenMP construct, and where to insert them in the 
sequential codes. Smart choice will generate efficient parallel codes,
while a bad choice of OpenMP directives may even generate a parallel code
with worse performance
than its original sequential code due to communication overheads.

When parallelize a loop in OpenMP,
we may also use the schedule clause to perform different scheduling policies
to effect how loop iterations are mapped onto threads.
There are four scheduling policies available in OpenMP.
The static scheduling method
deal-out blocks of iterations of siz ``chunk'' to each thread. 
In the dynamic scheduling method,
each thread grabs ``chunk'' iterations off a queue until all iterations have been handled. 
In the guided scheduling policy,
threads dynamically grab blocks of iterations. The size of the block starts large and 
shrinks down to siz ``chunk'' as the calculation proceeds.
Finally, in runtime scheduling method,
Schedule  and chunk size are taken from the {\tt OMP\_SCHEDULE} environment variable
and hence are determined at runtime.

The sections work-sharing construct gives a different structured block to each thread.  
This way, task parallelism can be implemented easily
if each section has a task (procedure call).
The following code shows that three tasks are parallelized using the
OpenMP section work-sharing
construct.

\begin{verbatim}
#pragma omp parallel
#pragma omp sections
{
        task1();
#pragma omp section
        task2();
#pragma omp section
        task3();
}
\end{verbatim}

Another important clause is the reduction clause,
that effects the way variables are shared.
The format is 
{\tt reduction (op $:$ list)},
where {\tt op} can be any general operations such as $+$, $max$, etc.
The variables in each ``list'' must be shared in the enclosing parallel region.  
Local copies are reduced into a single global copy at the end of the construct.
For example, here is an example for global sum and the final result is stored
in the variable $res$.

\begin{verbatim}
#include <omp.h> 
#define NUM_THREADS 2 
void main () 
{       int i;           
        double ZZ, func(), res=0.0;        
        omp_set_num_threads(NUM_THREADS) 
#pragma omp parallel for reduction(+:res) private(ZZ)        
        for (i=0; i< 1000; i++)
        {   
                ZZ = func(I);   
                res = res + ZZ;        
        } 
}
\end{verbatim}

Programming in a shared memory environment is generally easier
than in a distributed memory environment and thus saves labor costs. 
However, programming using message passing in distributed memory environment
usually produces more efficient parallel codes.
This is much like the relationship between
assembly languages and high level languages. 
Assembly codes run usually faster and are more compact than codes produced by
high-level programming languages and are used often in real-time or embedded
systems where both time and memory space are limited
and labor costs are not the primary consideration.
Besides producing efficient codes,
assembly languages are also useful when students learn basic concepts about
computer organization, arithmtic operation, machine languages,
addressing, instruction cycles, etc.
When we need to implement a large complicated program, high-level languages such as C,
C++, or Java are more frequently used.
Similarly, students can learn a lot of concepts such as scalability, 
broadcast, one-to-one communication, performance, communication overhead,
speedup, etc, through low-level languages such as MPI or PVM.
These concepts are hard to obtain through high-level parallel programming
languages due to the fact that many details are hidden in the language constructs.
However, students can implement a relatively large parallel program using
high-level parallel language such as OpenMP or HPF easily within a short period
of time.
We believe that the future of high performance computing
heavily depends on high level parallel programming languages
such as OpenMP due to the increasingly high labor costs and
the scarcity of good parallel programmers.
High level parallel programming languages are
 one way to make parallel computer systems popular and available to
non-computer scientists and engineers.
Hence, teaching students how to use high level parallel programming languages
as well as low level message passing paradigm
is an important task for teaching parallel programming.

\section{Why OpenMP and MPI}

There are currently four major standards for programming parallel systems that were 
developed in open forums: High Performance Fortran (HPF) \cite{hpf}, OpenMP \cite{openmp}, 
PVM \cite{pvm} and MPI \cite{mpi}. 

HPF relies on advanced compiler technology to expedite the development of data-parallel 
programs \cite{hpf}. 
Thus, although it is based on Fortran, HPF is a new language, and hence requires 
the construction of new compilers. As a consequence each implementation of HPF is, 
to a great extent, hardware specific, and until recently there were very few complete 
HPF implementations. Furthermore most of the current implementations are proprietary 
and quite expensive. 
HPF has been 
written for the express purpose of writing data-parallel programs, and, as a consequence 
it is not well-suited for dealing with irregular data-structures or 
control-parallel programs. 

The Parallel Virtual Machine (PVM) system
uses the message-passing model to allow programmers to exploit 
distributed computing across a wide variety of computer types, including multiprocessor
systems \cite{pvm}. A key concept in PVM is that it makes a collection of computers 
appear as one large virtual machine, hence its name. 
The PVM computing model is simple yet very general, and accommodates a wide variety 
of application program structures. The programming interface is 
deliberately straightforward, thus permitting simple program structures to be 
implemented in an intuitive manner. 
The user writes his application as a collection of cooperating tasks. 
Tasks access PVM resources through a library of standard interface routines. 
These routines allow the initiation and termination of tasks across the network 
as well as communication and synchronization between tasks. 
The PVM message-passing primitives are oriented towards heterogeneous operation, 
involving strongly typed constructs for buffering and transmission. 
Communication constructs include those for sending and receiving data structures 
as well as high-level primitives such as broadcast, barrier synchronization, and global sum.

MPI  specifies a library of extensions to C and Fortran that can be used to write 
message passing programs \cite{mpi}. 
So an implementation of MPI can make use of existing compilers, 
and it is possible to develop more-or-less portable MPI libraries. Thus, unlike HPF, 
it is relatively easy to find an MPI library that will run on existing hardware. 
All of these implementations can be freely downloaded from the internet.
Message passing is a completely general method for parallel programming. 
Indeed, the generality and ready availability of MPI 
have made it one of the most widely used systems for parallel programming. 
Compared with PVM library, MPI has become more popular recently.

Ideally, a parallel computing class would introduce students to both shared memory
and distributed memory programming models. However, 
it is a truism that learning to program parallel systems is extremely difficult, and we believe that attempting to learn two essentially different methodologies in one semester or one quarter will, especially for the weaker students, make it virtually impossible to learn either. Thus in view of the difficulties involved in obtaining an HPF compiler and the generality of MPI, we have chosen to use MPI.

OpenMP has emerged as the standard for shared memory parallel programming. 
For the first time, it is possible to write parallel programs 
which are portable across the majority of shared memory parallel computers. 
 OpenMP is a portable, scalable model that gives shared-memory parallel
programmers a simple and flexible interface for developing
parallel applications for platforms ranging from the desktop to
the supercomputer. 

The most important reason for us to adopt OpenMP in a parallel programming
class is that students can parallelize some realistic code (not toy problems)
within a short period of time due the programming ease it possesses.
Students can also experiment with different scheduling schemes such as static
 or dynamic loop scheduling policies within a short period of time,
which would be impossible otherwise using MPI.

Using OpenMP also has the advantage that task parallelism can be easily
implemented by just inserting several OpenMP directives.
By combining loop parallelism and task parallelism, better performance and
higher scalability can be achieved.
On the otehr hand, task parallelism is difficult to implement
using MPI or HPF.

Another reason that we selected OpenMP in our class is that
we have an SGI Origin 2000 shared memory multiprocessor system in our department.
A shared memory programming model will fit in well.

Besides the above reasons,
 OpenMP has the following benefits for parallel programming
compared with message passing models such as MPI:

a) A user just needs to add some directives into the sequential code to
instruct the compiler how to parallelize the code.
Hence, it has unprecedented programming ease,
making threading faster and more cost-effective than ever before.

b) The directives are treated as comments when running a single processor.
Hence, a single-source solution can be used for both
 serial and threaded applications,
lowering code maintenance costs.

c) It is portable parallelism across Windows NT and Unix platforms.

d) The correctness of the results generated using OpenMP
can be verified easily, to dramatically lower development and debugging costs.

Hence, our strategy is to teach students the basic concepts in parallel
programming such as scalability,                          
broadcast, one-to-one communication, performance, communication oevrhead,
speedup, etc, through a low-level parallel programming language, and
teach other concepts such as various scheduling policies and task parallelism
through a high-level parallel programming language.
Since MPI and OpenMP are the most widely used languages
in the two categories, we select them to teach parallel programming.

\section{Some Pitfalls with OpenMP}

Because OpenMP is a high level parallel language, many details are hidden from
a programmer. The good thing is that students can learn quickly and start to program
immediately after learning some techniques.
The pittfall is that students cannot clearly see the communications involved
in a parallel program. Our approach to overcome this problem
is to supplement OpenMP projects with some simple MPI programs.
Students first learn the basics of parallel programs in a distributed
memory environment. They start to parallelize a sequential code
using simple MPI constructs such as {\tt MPI\_Bcast}, 
{\tt MPI\_Reduce}, {\tt MPI\_Send}, and {\tt MPI\_Recv}.
Through several small projects, they learn the concepts of
one-to-one communication, multicast, broadcast, reduction, synchronization,
and concurrency.
Later, when they use OpenMP to parallelize a program,
they already have a deep understanding of communication structure,
communication overhead, scalability and performance issues.

The second shortcoming with OpenMP is that it does not provide memory
allocation schemes for arrays and other data structures since
OpenMP is designed for shared memory machines.
Again, this relieves the students from complication memory allocation
decisions and they can concentrate on loop and task parallelism.
This is good for the ease of programming, but students
do not know the details of array allocation schemes such as
BLOCK or CYCLIC schemes commonly used in distributed memory environments.
Since the memory on the SGI Origin 2000 is not physically shared, SGI
provides data distribution directives to allow users to specify how data
is placed on processors.
If no data distribution directives are used,
then data are automatically distributed via the ``first touch'' mechanism
\cite{sgi} which places the data on the processor where it is first used.
Because different allocation schemes may affect the performance
of a program greatly, SGI data distribution directives
are required in the final project to show the performance improvement.
For example, the following data distribution directive distribute the 4D array
H on dimension 2:
 
\begin{verbatim}
!$sgi distribute_reshape H(*,BLOCK,*,*)
\end{verbatim}
\noindent

Students are required to try several data distribution schemes, to observe
the running times and to comment on the timing results as described below.
In this way, the relationship between memory allocation schemes and performance
is demonstrated.

Due to these pitfalls with OpenMP, students would not learn all
the concepts and the whole picture in parallel programming using OpenMP alone.
Our strategy is to supplement OpenMP with explanation on 
several typical MPI codes and small projects
using MPI standard.
Then, students experiment with various scheduling policies and complicated
parallelization methods in OpenMP.
In this way, students experience various parallel schemes and techniques
 in a short period of time.
This would be very hard to achieve if only MPI or PVM programming model
is used in teaching parallel programming due to the facts that it is
very time consuming to implement in MPI or PVM and 
that students cannot parallelize a large code in MPI or PVM within a short period
of time.
The following section will detail the strategy of using both OpenMP and MPI
 adopted in my class.

\section{Using MPI and OpenMP in Projects}

The parallel programming class at GSU is a semester-long class for 
upper-level undergraduates and beginning graduate students. 
In it we used tutorials on MPI and OpenMP from Ohio Supercomputing Center 
as  supplements
to a parallel algorithms textbook  \cite{quinn}. 
The code that we present in the lectures uses both C and Fortran. 

The course begins with an overview of parallel computing and continues with a brief 
introduction to parallel computing models such as various PRAMs, shared memory
models and distributed memory models.
The concepts of data parallelism and pipelining are also introduced at that time.
The next block of lectures forms a transition into a more or less standard 
parallel algorithms course. We first discuss serial and parallel versions of a very 
simple computation -- e.g., prefix sums and prime finding. 
In the course of analyzing the performance of these algorithms, we develop the 
concepts of speedup, scalability and efficiency. 
The deterioration of the performance of the parallel algorithm as the number of 
processes is increased leads naturally into a discussion of Amdahl's Law and scalability. 

The course work consists of two tests, a final exam, five programming projects
and a research paper. 
Since the course's emphasis is on parallel programming,
projects are an important part of the course.
The purpose of the first project is simply to acquaint students with the system and
programming environment of the Origin 2000.
In it, they write a simple addition code, and measure the parallel times
using different number of processors.

In the second project, the students implement an MPI code to calculate $\pi$
using Simpson's Rule instead of the rectangle rule discussed in class,
where students are exposed to various MPI communication functions.
For timing measurements and precision,
they need to test the code using several different numbers of subintervals
to see the effect on the precision of results
and different number of processors on the execution times.

In the third project, the students implement parallel game of life. 
Through the assignment, students learn various domain decomposition strategies.
All the above projects are implemented in MPI.

Now students already understand the communication mechanism of
parallel computing systems, and communication overhaed within a parallel code,
it is time to introduce OpenMP.
After briefly discussing how to use OpenMP and showing them several OpenMP examples,
students are asked in the fourth project
to initialize a huge array $A$ so that each element has its index 
as its value. Then they are required to
create a real array $B$ which will contain the running average of 
array $A$. 
They need to parallelize the loops with all four scheduling schemes available in OpenMP
(static, dynamic, guided, and runtime)
and  measure the running times with different scheduling policies and different chunk sizes. 
They are also required to write their observations on the timings 
using the four different scheduling policies and to explain why the performance is different 
for using different scheduling policies and different chunk sizes. 

In the fifth project,
students learn how to parallel a real research Fortran code in OpenMP.
The project contains several parts.
They first parallelize the major loops                  
in the code in OpenMP.
Then, they parallelize the code using both loop and task parallelism in OpenMP.
Now, they have obtained the best scheduling policy, best chunk size for the policy,         
and both loop and task parallelism in the above two steps.
But array mapping is done automatically by the OpenMP compiler.
In the final step, they need to distribute the arrays manually using 
SGI array distribution directives since array distribution directives are
not available in OpenMP.
The purpose is for students to understand the effect of array distribution
on the running performance.
They are also required to write a short report                          
to summarize the results obtained.
Through these steps, students learned how to parallelize a real code
in a step-by-step fashion.

Students are also required to write a research paper or a survey paper
on a chosen topic in parallel processing.
The purpose is for the students to apply the knowledge learned in the course
to some problems.
Some students implemented some algorithms using MPI and/or OpenMP using various
strategies and compared the performance their implementations with the results published
in the literature.
At the end of the term, students need to present their findings besides submitting
a paper.

The outcome of the course is very good.
Most students feel that they learned a lot in the course
based on student evaluations and comments from them on the course.
Some of the students already applied their knowledge learned in the course
to our research projects supported by NSF and Air Force.
One student implemented a parallel program for Cholesky
factorization using both MPI and OpenMP 
and did a lot of testing using various scheduling and partition strategies.
He also did a comprehensive compariosn among the different implementations,
and wrote an excellent research paper at the end of the course.
The paper is being revised and potentially could be published in a conference.
This would have been impossible if only MPI had been taught in the course.

\section{Conclusion}

As OpenMP becomes more popular for parallel programming and has many advantages
over message passing programming models,
it is important to introduce OpenMP in a parallel programming course.
However, OpenMP also has some shortcomings for teaching parallel programming concepts.
Our strategy is to use MPI to convey the basic concepts of parallel programming
and to use OpenMP to tackle more complicated problems
such as various scheduling policies and combined loop and task parallelism.
It seems the strategy is well received by my students.

\end{document}